\documentclass[12pt]{article}

\usepackage{amsfonts}
  \usepackage{amsthm}
 \usepackage{amsmath}
 \usepackage{amscd}
 \usepackage{amssymb}
 \usepackage{latexsym}
\usepackage{mathdots}

 \numberwithin{equation}{section}
 \newtheorem{thm}{Theorem}
 
 \newtheorem{Cor}{Corollary}

 \newtheorem{prop}{Proposition}
 \theoremstyle{definition}

\title{Near-integrability  of periodic Klein-Gordon lattices}
\author{Ognyan Christov \\
Faculty of Mathematics and Informatics, Sofia University, \\
5 J. Bourchier blvd. 1164 Sofia, Bulgaria}

\date{}

\begin{document}

\maketitle

\begin{abstract}
In this paper we study the Klein-Gordon (KG) lattice with periodic
boundary conditions. It is an $N$ degrees of  freedom Hamiltonian
system with linear inter-site forces and nonlinear on-site
potential, which here is taken to be of the $\phi^4$ form.

First, we prove that the system in consideration  is
non-integrable in Liuville sense. The proof is based on the
Morales-Ramis theory.

Next, we deal with the resonant Birkhoff normal form of the KG
Hamiltonian, truncated to order four. Due to the choice of
potential, the periodic KG lattice shares the same set of discrete
symmetries as the periodic Fermi-Pasta-Ulam (FPU) chain. Then we
show that the above normal form is integrable. To do this we
utilize the results of B. Rink on FPU chains.

If $N$ is odd this integrable normal form turns out to be  KAM
nondegenerate Hamiltonian. This implies the existence of many
invariant tori at low-energy level in the dynamics of the periodic
KG lattice, on which the motion is quasi-periodic.

We also prove that the KG lattice with Dirichlet boundary
conditions (that is, with fixed endpoints) admits an integrable,
KAM nondegenerated normal forth order form, which in turn shows
that almost all low-energetic solutions of KG lattice with fixed
endpoints are quasi-periodic.

\end{abstract}

\section{Introduction}

Let us introduce the Klein-Gordon (KG) lattice described by the
Hamiltonian
\begin{equation}
\label{1.1} H = \sum_{j \in \mathbb{Z}} \Big[\frac{p_j^2}{2} +
\frac{C}{2}(q_{j+1} - q_j)^2 + V (q_j) \Big], \quad p_j =
\dot{q}_j .
\end{equation}
The constant $C > 0$ measures the interaction to nearest neighbor
particles (with unit masses) and $V (x)$ is a non-linear
potential. This lattice appears, for instance, as a spatial
discretization of the Klein-Gordon equation
\begin{equation}
\label{KGeq} u_{tt} = u_{xx} +  V (u).
\end{equation}
Both models subjected to different boundary conditions are used to
describe a wide variety of physical phenomena: crystal
dislocation, localized excitations in ionic crystals (see e.g.
\cite{Morgante}, \cite{IoossPel}). In particular, the model
(\ref{1.1}) with the Morse potential $V (x) = D (e^{-\alpha x}
-1)^2$ is applied to studying the  thermal denaturation of DNA
\cite{Peyrard}.

Our aim is to study the regular behavior of the trajectories of
(\ref{1.1}) and in particular, to study the complete integrability
of the corresponding Hamiltonian system. When $C=0$ the
Hamiltonian is separable, and hence, integrable. There exist
plenty of periodic or quasi-periodic solutions in the dynamics of
(\ref{1.1}). It is natural to investigate whether this behavior
persists for $C$ small enough (see e.g. \cite{Jong}). At this
point it is worth mentioning that in {\it anticontinuous} limit $C
\to 0$, there exist self-localized periodic oscillations  called
also {\it discrete breathers} \cite{MacAu}, \cite{FlWi}). Here we
assume that $C$ is neither very small nor too large and put  $C =
1$, which can be achieved by rescaling  $t$.

We are interested mainly in the behavior at low energy, so we take
quartic ($\phi^4$) potential
\begin{equation}
  \label{1.11}
   V (x) = \frac{a}{2} x^2 + \frac{\beta}{4} x^4,   \qquad    a > 0 ,
\end{equation}
which is  frequently used in the research on the subject (see
\cite{Morgante} and the literature therein).

First, periodic boundary conditions for (\ref{1.1}) are assumed.
Then one gets system with $N$ degrees of freedom,  described by
the Hamiltonian
\begin{equation}
\label{1.2} H = \sum_{j \in \mathbb{Z}/N\mathbb{Z}}
\Big[\frac{p_j^2}{2} + \frac{1}{2}(q_{j+1} - q_j)^2 +
\frac{a}{2}(q_j)^2 + \frac{\beta}{4} (q_j)^4 \Big], \quad p_j =
\dot{q}_j .
\end{equation}
Our first result concerns the integrability of the Hamiltonian
system governed by (\ref{1.2}).
\begin{thm}
\label{th1} The periodic KG lattice with Hamiltonian (\ref{1.2})
is integrable only when $\beta = 0$.
\end{thm}

\vspace{3ex}

Motivated by the works of Rink \cite{RinkVer,Rink1}, who presented
the periodic FPU chain as a perturbation of an integrable and KAM
non-degenerated system, namely the truncated Birkhoff-Gustavson
normal form of order 4 in the neighborhood of an equilibrium, we
would like to verify whether this can be done for KG lattices.

As in the case of periodic FPU chain the properties of the
periodic KG lattice near the equilibrium strongly depend on the
parity of the number of the particles $N$. Assume, in addition,
that $a = 1$ (see an explanation for this choice in the next
section). Then we have the following
\begin{thm}
\label{th2} The fourth order  normal form $\overline{H} = H_2 +
\overline{H}_4$ of the periodic KG lattice is:

   (i) completely integrable and KAM non-degenerate for $N$ odd;

   (ii) completely integrable for $N$ even.
\end{thm}
{\bf Remark 1.} The statement of Theorem \ref{th2} will be made
more precise in section 3. Note that the cases with $N=2$ and
$N=4$ particles are rather exceptions to the general situation.
The corresponding first integrals are quadratic and the KAM
conditions are trivially checked.

\vspace{3ex}

As a consequence from this result, we may conclude for the
periodic KG lattices when KAM theory applies, that there exist
many quasi-periodic solutions of small energy on a long time scale
and chaotic orbits are of small measure.

Next, Dirichlet boundary conditions for (\ref{1.1}) are
considered. Due to Rink \cite{Rink3} in the case of FPU chain such
a system can be viewed as an invariant symplectic submanifold of a
periodic FPU chain. This approach also works for KG lattice and
the corresponding result is as follows.
\begin{thm}
\label{th3} The fourth order  normal form $\overline{H} = H_2 +
\overline{H}_4$ of KG lattice with fixed endpoints is completely
integrable and KAM nondegenerate.
\end{thm}

This result and KAM theorem show the existence of large-measure
set of low-energy quasi-periodic solutions of KG lattice with
fixed endpoints.

\vspace{3ex}

\noindent {\it Related works:} Recall that the FPU chain with $n$
particles (with unit masses) and with periodic boundary conditions
can be described by a Hamiltonian system with the Hamiltonian
\begin{equation}
\label{fpu} H = \sum_j= \frac{1}{2} p_j ^2 + W (q_{j+1} - q_j),
\quad p_j = \dot{q}_j, \qquad (q_0, p_0) = (q_n, p_n) ,
\end{equation}
where $W$ is a real potential of the kind $ W (x) = \frac{1}{2!}
x^2 + \frac{\alpha}{3!} x^3 + \frac{\beta}{4!} x^4 . $ When
$\alpha \neq 0, \beta = 0$ the chain is called an $\alpha$-chain.
Accordingly, when $\alpha = 0, \beta \neq 0$ the chain is known as
a $\beta$-chain.

Probably T. Nishida  \cite{Nishida} and J. Sanders \cite{Sanders}
are among the first who have calculated the normal form of the FPU
chain with fixed endpoints and with periodic boundary conditions,
respectively. By imposing some very strong non-resonant
assumptions, they verify the KAM theory conditions, but in general
resonances do exist.

We already mentioned the works of Rink who has proved rigorously
that the periodic FPU Hamiltonian is a perturbation of a
nondegenerate Liouville integrable Hamiltonian, namely the normal
form of order 4, \cite{Rink1}. Furthermore, he has described the
geometry of even FPU lattice in \cite{Rink2}, and finally
rigorously has proved the Nishida's conjecture stating that almost
all low-energetic motions in FPU with fixed endpoints are
quasi-periodic \cite{Rink3}.

One should note that the Rink's results are consequence of the
special symmetry and resonance properties of the FPU chain and
should not be expected for lower-order resonant Hamiltonian
systems (see e.g. \cite{OC1}).

However, several  open problems remain: some of them purely
computational and some of them of more philosophical nature.
Henrici and Kappeler  \cite{HK1,HK2} managed to solve practically
all these problems, and generalized the results of Rink, by
applying special sets of canonical variables, originally were
designed for the Toda chain.

\vspace{3ex}

Our study on the normal forms of the KG lattice (in particular,
Theorems \ref{th2} and \ref{th3}) is related to the results of
Rink on the normal form of the periodic FPU chain and use his
approach. Notice that there are some differences between the
potentials defining the models describing FPU and KG lattices.
Moreover, the periodic FPU chain with three particles is
integrable while Theorem \ref{th1} shows that this is not the case
for the periodic KG lattice.

Our goal is to see whether these differences affect the
integrability and eventually the dynamics of the system
corresponding to the truncated normal form. In view of the wide
applicability of the KG models, we think that this study is
naturally motivated.

\vspace{3ex}

\noindent The paper is organized as follows. In section 2 we
recall some known facts about Liouville integrable systems,
action-angle variables, KAM conditions and normal forms. We also
consider the resonances and discrete symmetries of the considered
Hamiltonian system. In section 3 we give the truncated to order
four normal form for the periodic KG lattice and consider the
integrability of this normal form for $N$ odd and even,
respectively. For any of the cases a more detailed description of
the commuting first integrals is given. This proves Theorem
\ref{th2}. Section 4 is devoted to the KG lattice with fixed
endpoints. Making use of  the result from the previous section,
the truncated normal form of the corresponding Hamiltonian is
derived and  Theorem \ref{th3} is proved. We finish with some
remarks and possible directions to extend our results.

The proof of Theorem \ref{th1} is based on the Morales-Ramis
theory and since it is more algebraic in nature, it is carried out
in the Appendix.

\section{Resonances and symmetries}

In this section we recall briefly some notions and facts about
integrability of Hamiltonian systems, action-angle variables,
perturbation of integrable systems and normal forms. More complete
exposition can be found in \cite{AKN}.

 Let $H$ be an analytic Hamiltonian defined on a $2 n$ dimensional
symplectic manifold. The corresponding
 Hamiltonian system is
\begin{equation}
\label{2.1}
 \dot x = X_H (x).
 \end{equation}
It is said that a Hamiltonian system is completely integrable if
there exist $n$ independent integrals $F_1 = H, F_2, \ldots, F_n$
in involution, namely $\{ F_i, F_j \} = 0$ for all $i$ and $j$,
where $\{ , \}$ is the Poisson bracket. On a neighborhood $U$ of
the connected compact level sets of the integrals $M_c = \{ F_j =
c_j, j = 1, \ldots, n \}$  by Liouville - Arnold theorem one can
introduce a special set of symplectic coordinates, $ I_j,
\varphi_j$, called action - angle variables. Then, the integrals
$F_1 = H, F_2, \ldots, F_n$  are  functions of action variables
only and the flow of $X_H$ is simple
\begin{equation}
\label{2.2} \dot{I}_j = 0, \quad \dot{\varphi}_j = \frac{\partial
H}{\partial I_j}, \quad j = 1, \ldots, n.
\end{equation}
Therefore, near $M_c$, the phase space is foliated with $X_{F_i}$
invariant tori over which the flow of $X_H$ is quasi - periodic
with frequencies $ (\omega_1 (I), \ldots, \omega_n (I) ) = (
\frac{\partial H}{\partial I_1}, \ldots, \frac{\partial
H}{\partial I_n})$.

The map
\begin{equation}
\label{2.3} (I_1, I_2, \ldots, I_n) \to \left( \frac{\partial
H}{\partial I_1}, \frac{\partial H}{\partial I_2}, \ldots,
\frac{\partial H}{\partial I_n}\right)
\end{equation}
is called frequency map.

Consider a small perturbation of an integrable Hamiltonian $H_0
(I)$. According to Poincar\'{e} the main problem of mechanics is
to study the perturbation of quasi-periodic motions in the system
given by the Hamiltonian
$$
H = H_0 (I) + \varepsilon H_1 (I, \varphi), \quad \varepsilon < <
1.
$$
 KAM - theory  gives conditions on the integrable Hamiltonian $H_0$ which ensures
the survival of the most of the invariant tori. The following
condition, usually called Kolmogorov's condition, is that the
frequency map should be a local diffeomorphism, or equivalently
\begin{equation}
\label{2.4} \det\left( \frac{\partial^2 H_0}{\partial I_i \partial
I_j}  \right) \neq 0
\end{equation}
on an open and dense  subset of $U$. We should note that the
measure of the surviving tori decreases with the increase of both
perturbation and the measure of the set where above Hessian is too
close to zero.


 In a neighborhood of the equilibrium $(q, p) = (0, 0)$ we have the
 following expansion of $H$
\begin{align*}
H   & = H_2 + H_3 + H_4 + \ldots, \\
H_2 & = \sum \omega_j ( q_j ^2 + p_j ^2 ), \quad \omega_j > 0 .
\end{align*}
We assume that $H_2$ is a positively defined quadratic form. The
frequency $\omega = (\omega_1, \ldots, \omega_n)$ is said to be in
resonance if there exists a vector $k = (k_1, \ldots, k_n), \, k_j
\in \mathbb{Z}, j = 1, \ldots, n$, such that $(\omega, k) = \sum
k_j \omega_j = 0$,  where $|\, k | = \sum |\, k_j |$  is the order
of resonance.

 With the help of a series of canonical transformations
close to the identity, $H$  simplifies. In the absence of
resonances the simplified Hamiltonian is called  Birkhoff normal
form, otherwise -  Birkhoff-Gustavson normal form, which may
contain combinations of angles arising from resonances.

Often to detect the behavior in a small neighborhood of the
equilibrium, instead of the Hamiltonian $H$ one considers the
normal form truncated to some order
$$
\overline{H} = H_2 + \ldots + \overline{H}_m .
$$
It is known that the truncated to any order Birkhoff normal form
is integrable \cite{AKN}. The truncated Birkhoff-Gustavson normal
form has at least two integrals - $H_2$ and $\overline{H}$.
Therefore, the truncated resonant normal form of two degrees of
freedom Hamiltonian is integrable.

In order to obtain estimates of the approximation by normalization
in a neighborhood of an equilibrium point we scale $q \to
\varepsilon \tilde{q}, \, p \to \varepsilon \tilde{p}.$ Here
$\varepsilon$ is a small positive parameter and $\varepsilon^2$ is
a measure for the energy relative to the equilibrium energy. Then,
dividing by $\varepsilon^2$ and removing tildes we get
$$
 \overline{H} = H_2 + \varepsilon \overline{H}_3 + \ldots + \varepsilon^{m-2} \overline{H}_m .
$$
 Provided that $\omega_j > 0$ it is proven  in \cite{V1} that $\bar{H}$
is an integral for the original system with error $O
(\varepsilon^{m-1})$
 and $H_2$ is an integral for the original system with error $O (\varepsilon)$ for  the whole time interval.
 If we have more  independent integrals, then they are integrals
 for the original Hamiltonian system with error $O (\varepsilon^{m-2})$
 on the time scale $1/\varepsilon$.

 The first integrals for the normal form  $\overline{H}$ are approximate integrals for the original system,
 that is,   if the normal form is integrable then the original system is
 {\bf near integrable}  in the above sense.


Returning to the Hamiltonian of the periodic KG lattice
(\ref{1.2}) we see that its  quadratic part $H_2$  is not in
diagonal form
\begin{equation}
\label{2.5} H_2 = \frac{1}{2} p^T p + \frac{1}{2} q^T L_N q.
\end{equation}
Here $L_N$ is the following $N \times N$ matrix
\begin{equation}
\label{2.6} L_N :=
\begin{pmatrix}
2+a & -1 &   &   & -1 \\
-1 & 2+a & -1  &   &  \\
\\
   & \ddots & \ddots   & \ddots       &  \\
   \\
  &   &-1   &  2+a & -1   \\
-1 &   &   & -1 & 2+a
\end{pmatrix} .
\end{equation}
The eigenvalues of $L_N$ are of the form $\omega_k ^2 = a +  4
\sin^2 \frac{k \pi}{N}$. There is a symplectic
Fourier-transformation $q = M Q, p = M P$ which brings $H_2$ in
diagonal form
\begin{equation}
\label{2.10} H_2 = \frac{1}{2} P^T P + \frac{1}{2} Q^T \Omega Q,
\end{equation}
where $M ^{-1} L_N M = \Omega := \mathrm{diag} (\omega_1 ^2,
\ldots, \omega_N ^2)$. The variables $(Q, P)$ are known as {\bf
phonons}. Denote for short
$$
 c_{kj} : =  \cos \left(\frac{2 k j \pi}{N} \right),
\quad s_{kj} : =  \sin \left(\frac{2 k j \pi}{N} \right).
$$
Then the transform $(q, p) \to (Q, P)$ in coordinate form is
\begin{eqnarray}
\label{prava}
q_j = & \sqrt{\frac{2}{N}} \sum\limits_{1\leq k < \frac{N}{2}} c_{kj} Q_k + s_{kj} Q_{N-k} + \frac{(-1)^j}{\sqrt{N}} Q_{\frac{N}{2}} + \frac{1}{\sqrt{N}} Q_N\\
p_j = & \sqrt{\frac{2}{N}} \sum\limits_{1\leq k < \frac{N}{2}}
c_{kj} P_k + s_{kj} P_{N-k} + \frac{(-1)^j}{\sqrt{N}}
P_{\frac{N}{2}} + \frac{1}{\sqrt{N}} P_N\nonumber
\end{eqnarray}
From these formulas one can easily get the explicit form of the
matrix $M$. Later on, we will make use of the inverse transform:
for $1 \leq j < \frac{N}{2}$ we have
\begin{eqnarray}
\label{obr}
Q_j &=& \sqrt{\frac{2}{N}} \sum_{k=1} ^N c_{kj} q_k, \qquad P_j = \sqrt{\frac{2}{N}} \sum_{k=1} ^N c_{kj} p_k, \nonumber \\
Q_{N-j} &=& \sqrt{\frac{2}{N}} \sum_{k=1} ^N s_{kj} q_k, \qquad P_{N-j} = \sqrt{\frac{2}{N}} \sum_{k=1} ^N s_{kj} p_k, \\
Q_N &=& \frac{1}{\sqrt{N}} \sum_{k=1} ^N q_k, \qquad P_N =
\frac{1}{\sqrt{N}} \sum_{k=1} ^N p_k \nonumber
\end{eqnarray}
and if $N$ is even
$$
Q_{\frac{N}{2}} = \frac{1}{\sqrt{N}} \sum_{k=1} ^N (-1)^k q_k,
\qquad P_{\frac{N}{2}} = \frac{1}{\sqrt{N}} \sum_{k=1} ^N (-1)^k
p_k.
$$
Finally, by simple scaling $Q_k \to \frac{1}{\sqrt{\omega_k}} Q_k,
\, P_k \to \sqrt{\omega_k} P_k$  we get the desired form of $H_2$
\begin{equation}
\label{nH2} H_2 = \sum_{k=1} ^N \frac{\omega_k}{2} (P_k^2 + Q_k
^2).
\end{equation}

Next, we need to study the possible relations between the
frequencies $\omega_k$. In general,  $\omega_k = \omega_j$ holds
if only if $j = N - k$. The resonances $\omega_k : \omega_{N-k} =
(1 : 1)$ (known also as internal resonances \cite{RinkVer}) are
important for the construction of the normal form.

The assumption on $a$  does not prevent the appearance of more
complicated resonances. It is clear that
 there are plenty of resonances when $a \in \mathbb{Q}$.
Moreover, there are certain  irrational values of $a > 0$  for
which fourth order resonances exist in the low-dimensional
periodic KG lattices (see \cite{OC2}). Such values of $a$ are
difficult to control in the higher dimensions, that is why from
now on we put $a=1$
\begin{equation}
\label{2.11} \omega_k = \sqrt{1 + 4 \sin ^2 \frac{k \pi}{N}}  .
\end{equation}
Note that another resonant relation 2 : 2 : 1 appears when $N = 3
s$. The other possible fourth order relations are

1) $\omega_k = 3 \omega_{k'} $ \quad  \, \, \,  \quad 2) $\omega_k
=  \omega_{k'} + \omega_{k''}  + \omega_{k'''}$

3) $  \omega_k + \omega_{k'} = 2 \omega_{k''} $ \quad   4)  $
\omega_k  + \omega_{k'} =  \omega_{k''}  + \omega_{k'''} $

\noindent for some $1 \leq k, k', k'', k''' \leq N$. It is
straightforward that  there are no such $k, k', k'', k'''$ to
fulfill 1) and 2) since $1 \leq \omega_k \leq \sqrt{5}$. Since 3)
is a particular case of the last relation, let us consider 4).
Suppose the tuple $(k, k', k'', k''')$ is a solution of
$$
\sqrt{1 + 4 \sin ^2 \frac{k \pi}{N}} + \sqrt{1 + 4 \sin ^2
\frac{k' \pi}{N}} = \sqrt{1 + 4 \sin ^2 \frac{k'' \pi}{N}} +
\sqrt{1 + 4 \sin ^2 \frac{k''' \pi}{N}}.
$$
Then such is the tuple $(N-k, N-k', N-k'', N-k''')$ and any
combination between two of them. Then in searching for these
tuples we can consider only the cases where $ k + k' + k'' + k'''
\equiv 0 \mod{N}$.

\vspace{2ex}

{\bf Assertion.} The only possible tuples $(k, k', k'', k''')$,
which satisfies 4) are those that can be derived from $(k, k',
N-k, N-k')$ by permutations.

\vspace{2ex}

{\bf Remark 2.} So far we have no rigorous  proof for that claim,
but we believe so.
 Direct computations show that this assertion is true for low dimensional cases $N = 2, \ldots, 6$.
Numerical simulations are confirmative.

\vspace{2ex}

Finally, in order to keep symmetry in the formulas we continue to
write $\omega_N$ and $\omega_{N/2}$ instead of their particular
values $1$ and $\sqrt{5}$.

The Hamiltonian (\ref{1.2}) of the periodic KG lattice possesses
discrete symmetries. Two of them, important for the dynamics and
exactly the same as in the periodic FPU chain, are the linear
mappings $ R, S :T^*\mathbb{R}^N \to T^*\mathbb{R}^N$ defined by
(see \cite{Rink1, Rink3})
\begin{align}
\label{symT}
R : & (q_1, q_2, \ldots, q_{N-1}, q_N; p_1, p_2, \ldots, p_{N-1}, p_N)  \nonumber \\
    & \mapsto (q_2, q_3, \ldots, q_N, q_1; p_2, p_3, \ldots, p_N, p_1)
\end{align}
and
\begin{align}
\label{symS}
S : & (q_1, q_2, \ldots, q_{N-1}, q_N; p_1, p_2, \ldots, p_{N-1}, p_N) \nonumber \\
    &   \mapsto -(q_{N-1}, q_{N-2}, \ldots, q_1, q_N; p_{N-1}, p_{N-2}, \ldots, p_1, p_N).
\end{align}
It is easily seen that $R$ can serve as a generator of a group
$\langle R\rangle$, isomorphic to the cyclic group of order $N$
and $S$ as a generator of $\langle S\rangle$, isomorphic to the
cyclic group of order two. Note that $R$ and $S$ are canonical
transformations $R^* (d q \wedge d p) = S^* (d q \wedge d p) = d q
\wedge d p $ and they leave the Hamiltonian invariant $ R^* H = (H
\circ R) = S^* H = H $. Moreover, they leave the Hamiltonian
vector field $X_H$ invariant, which implies that they commute with
the flow of $X_H$. It is observed in \cite{Rink1} that the
subgroup $\langle R, S\rangle := \{Id, R, \ldots, R^{N-1}, SR,
\ldots, SR^{N-1} \}$ of the symmetry group of $H$, with the
relations $R^N =S^2 = Id, S R = R^{-1} S$, is isomorphic to the
$N$th dihedral group.

For Hamiltonian systems with symmetry, we have
\begin{thm}
\label{thsym} (see \cite{ChKR} and \cite{Gaeta}). Let $H = H_2 +
H_3 + \ldots$ be the expansion of $H$ in a neighborhood of
equilibrium and $G$ be a group of linear symplectic symmetries of
$H$. Then a normal form $\overline{H} = H_2 + \ldots +
\overline{H}_m $ for $H$ can be constructed in such a way that
$\overline{H}$ is also $G$-symmetric.
\end{thm}

\section{The Birkhoff normal form}

In phonon coordinates, the Hamiltonian (\ref{1.2}) is
\begin{equation}
\label{3.1} H =  \sum_{k=1} ^N \frac{1}{2} \omega_k (P_k^2 +  Q_k
^2) + H_4 (Q_1, \ldots, Q_N) .
\end{equation}
Further, we introduce the complex variables
$$
z_k = Q_k + i P_k, \quad \xi_k = Q_k - i P_k,
$$
which are not symplectic, but are natural in the construction of
the normal form. In these variables $H_2$ reads
\begin{equation}
\label{3.3} H_2 = \sum\limits_{1\leq k <\frac{N}{2}} \omega_k (z_k
\xi _k + z_{N-k} \xi_{N-k}) + \omega_{\frac{N}{2}} z_{\frac{N}{2}}
\xi_{\frac{N}{2}}
 + \omega_{N} z_{N} \xi_{N}.
\end{equation}
Next, we are looking for the monomials $z^{\Theta} \xi^\theta$,
$\Theta, \theta$ being multi-indices, which commute with $H_2$,
i.e. $ad_{H_2} (z^{\Theta} \xi^\theta) = 0$. These monomials are
called resonant monomials and cannot be removed in the process of
normalization. We then get
\begin{equation}
\label{3.4} ad_{H_2} (z^{\Theta} \xi^\theta) = \nu (\Theta,
\theta) z^{\Theta} \xi^\theta
\end{equation}
with
\begin{equation}
\label{3.5} \nu (\Theta, \theta) := \sum\limits_{1\leq k
<\frac{N}{2}} i \omega_k (\Theta_k - \theta_k + \Theta_{N-k} -
\theta_{N-k} ) + i \omega_{\frac{N}{2}} (\Theta_{\frac{N}{2}} -
\theta_{\frac{N}{2}}) + i \omega_{N} ( \Theta_N - \theta_N).
\end{equation}
Hence, the resonant monomials are ones with $\nu (\Theta, \theta)
= 0$. Therefore, modulo the Remark 2 from the previous section we
have that the set of multi-indices $(\Theta, \theta)$ for which $|
\Theta | + | \theta | = 4$ and $\nu (\Theta, \theta) = 0$ is
contained in the set given by the relations
$$
\Theta_k - \theta_k + \Theta_{N-k} - \theta_{N-k} =
\Theta_{\frac{N}{2}} - \theta_{\frac{N}{2}} =  \Theta_N - \theta_N
= 0
$$
This means that $\overline{H}_4$ is generated by
$$
z_k \xi_k, \, z_{N-k} \xi_{N-k}, z_k \xi_{N-k}, \, \xi_k z_{N-k},
\quad 1 \leq k < \frac{N}{2} \quad \mbox{and} \quad
z_{\frac{N}{2}} \xi_{\frac{N}{2}}, \, z_N \xi_N
$$

However, we want a normal form which is invariant under $R^*$ and
$S^*$. To obtain such we define for $1 \leq j < \frac{N}{2}$
\begin{align}
\label{Hopfv}
a_k &:= \frac{1}{2} (z_k \xi_k + z_{N-k} \xi_{N-k}) = \frac{1}{2} (Q_k ^2 + P_k ^2 + Q_{N-k} ^2 + P_{N-k} ^2),  \nonumber \\
b_k &:= \frac{i}{2} (z_k \xi_{N-k} - z_{N-k} \xi_k ) = Q_k P_{N-k} - Q_{N-k} P_k,    \nonumber \\
c_k &:= \frac{1}{2} (z_k \xi_k + z_{N-k} \xi_{N-k}) = \frac{1}{2} (Q_k ^2 + P_k ^2 - Q_{N-k} ^2 - P_{N-k} ^2), \\
d_k &:= \frac{1}{2} (z_k \xi_{N-k} + z_{N-k} \xi_k ) = Q_k Q_{N-k}
+ P_k P_{N-k}, \nonumber
\end{align}
$ a_N :=  \frac{1}{2} z_N \xi_N =  \frac{1}{2} (Q_N ^2 + P_N ^2) $
and if $N$ is even $a_{\frac{N}{2}} :=  \frac{1}{2}
z_{\frac{N}{2}} \xi_{\frac{N}{2}} =  \frac{1}{2} (Q_{\frac{N}{2}}
^2 + P_{\frac{N}{2}} ^2)$.

These quantities are known as Hopf variables and they satisfy the
relations
\begin{equation}
\label{relHopfv} a_k ^2 = b_k ^2 + c_k ^2 + d_k ^2, \quad 1 \leq k
< \frac{N}{2}
\end{equation}
and in these variables $H_2$ is
\begin{equation}
\label{h2} H_2 = \sum\limits_{1\leq k < \frac{N}{2}} \omega_k  a_k
+ \omega_{\frac{N}{2}} a_{\frac{N}{2}} + \omega_N a_N .
\end{equation}

The nontrivial Poisson brackets between these quantities are
\begin{equation}
\label{brackets} \{b_k, c_k\} = 2 d_k, \quad \{b_k, d_k\} = - 2
c_k, \quad \{c_k, d_k\} = 2 b_k.
\end{equation}

It is observed in \cite{Rink1} that $a_k$, $a_{\frac{N}{2}}$ and
$a_n$ are invariant under $R^*$ and $S^*$ and also the products
$a_k a_l, a_N a_k, b_k b_l, 1 \leq k, l < \frac{N}{2} $ and if $N$
is even $ a_{\frac{N}{2}} a_k, 1 \leq k \leq \frac{N}{2}$ and  $
d_k d_{\frac{N}{2}-k} - c_k c_{\frac{N}{2}-k}, 1 \leq k \leq
\frac{N}{4}$, so $\overline{H}_4$ must be a linear combination of
the above four order terms. Indeed, we have
\begin{thm}
\label{thNF} The truncated up to order four normal form for the
periodic KG lattice is
$$
\overline{H} = H_2  + \overline{H}_4 ,
$$
\begin{align}
\label{h4} \overline{H}_4 & = \frac{\beta}{2 N} \Bigg \{
\frac{3}{2} \left( \frac{a_{\frac{N}{2}} ^2}{\omega_{\frac{N}{2}}
^2} + \frac{a_N ^2}{\omega_N ^2} \right) + 6 \frac{a_{\frac{N}{2}}
a_N }{ \omega_{\frac{N}{2}} \omega_N } + 6 \left(
\frac{a_{\frac{N}{2}}}{ \omega_{\frac{N}{2}}} +
\frac{a_N}{\omega_N}\right) \sum\limits_{1\leq k < \frac{N}{2}}
\frac{a_k}{\omega_k}
+ \frac{3}{4} \sum\limits_{1\leq k < \frac{N}{2}} \frac{3a_k^2 -b_k^2}{\omega_k^2} \nonumber \\
& + 6 \sum\limits_{1\leq k < l < \frac{N}{2}} \frac{a_k
a_l}{\omega_k \omega_l} + 3 \sum\limits_{1\leq k < \frac{N}{4}}
\frac{c_k c_{\frac{N}{2}-k} - d_k d_{\frac{N}{2}-k}}{\omega_k
\omega_{\frac{N}{2}-k}} + \frac{3}{4} \frac{c_{\frac{N}{4}}^2 -
d_{\frac{N}{4}}^2}{\omega_{\frac{N}{4}}^2} \Bigg \} .
\end{align}
In the above formula the terms with subscripts $\frac{N}{2}$,
$\frac{N}{4}$ appear if $\frac{N}{2} \in \mathbb{N}$, $\frac{N}{4}
\in \mathbb{N}$, respectively.
\end{thm}
The calculation of the above normal form is long, tedious, but
straightforward, that's why it is not presented here. Instead, we
proceed with two important corollaries, which prove the assertions
in Theorem \ref{th2}.
\begin{Cor}
\label{Cor1} When $N$ is odd, the truncated normal form
$\overline{H} = H_2 + \overline{H}_4$ of the periodic KG lattice
is Liouville integrable with the quadratic integrals $a_j, b_j, \,
1 \leq j \leq \frac{N-1}{2}$ and $a_N$. Moreover, this normal form
is KAM nondegenerate.
\end{Cor}
\noindent {\bf Proof.} $H_2$ is a linear combination of $a_j$ and
$a_N$. When $N$ is odd, $\overline{H}_4$ becomes
\begin{align}
\label{3.10} \overline{H}_4 & = \frac{\beta}{2N} \Bigg
\{\frac{3}{4} \sum_{k=1} ^{(N-1)/2} \frac{3a_k^2
-b_k^2}{\omega_k^2} + \frac{3}{2} \frac{a_N ^2}{\omega_N ^2} + 6
\frac{a_N}{\omega_N} \sum_{k=1} ^{(N-1)/2} \frac{a_k}{\omega_k} +
6 \sum\limits_{1\leq k < l \leq \frac{N-1}{2}} \frac{a_k
a_l}{\omega_k \omega_l} \Bigg \}
\end{align}
and it is clear that $a_j, b_j, \, 1 \leq j \leq \frac{N-1}{2}$
and $a_N$ commute with $\overline{H}_4$ and with each other.

In order to introduce action-angle variables we follow the scheme
from  \cite{Rink1}, slightly adjusted to our case. We need to find
the set of regular values of the energy momentum map
$$
EM : (Q, P) \to (a_j, b_j, a_N).
$$
Denote it by $U_r = \{ (a_j, b_j, a_N) \in \mathbb{R}^N, \, a_j >
0 , | b_j | < a_j, \, a_N > 0 \}$. Then for all $(a_j, b_j, a_N)
\in U_r$ the level sets of $EM^{-1} (a_j, b_j, a_N)$ are
diffeomorphic to N-tori.

Let $\mathrm{arg}: \mathbb{R}^2 \setminus \{(0, 0)\} \to
\mathbb{R}/2 \pi \mathbb{Z}$ be the argument function
$\mathrm{arg} (r \cos \Phi, r \sin \Phi) \to \Phi$. Define the
following set of variables $ (a_j, b_j, a_N, \phi_j, \psi_j,
\varphi_N)$  $a_j, b_j, a_N$ as above and
\begin{align}
\label{3.6} \phi_j := & \, \frac{1}{2} \mathrm{arg}(-P_{N-j} -
Q_j, P_j - Q_{N-j}) +
\frac{1}{2} \mathrm{arg}(P_{N-j} - Q_j, P_j + Q_{N-j}), \nonumber \\
\psi_j := & \,  \frac{1}{2} \mathrm{arg}(-P_{N-j} - Q_j, P_j -
Q_{N-j}) -
\frac{1}{2} \mathrm{arg}(P_{N-j} - Q_j, P_j + Q_{N-j}),  \\
 \varphi_N : = & \, \arctan P_N/ Q_N . \nonumber
\end{align}
Using the formula $d \mathrm{arg} (x, y) = \frac{x d y - y dx}{x^2
+ y^2}$, one can verify that $(a_j, b_j, a_N, \phi_j, \psi_j,
\varphi_N)$ are indeed canonical coordinates $\sum d P_j \wedge d
Q_j = \sum d a_j \wedge d \phi_j + d b_j \wedge d \psi_j + d a_N
\wedge d \varphi_N$. Since $a_j, b_j$ and $a_N$ are quadratic
functions in the phase variables, they can be extended to global
action variables.

Finally to check the nondegeneracy condition, we compute the
Hessian of $\overline{H}_4$ with respect to $a_j, a_N, b_k$.
Denote $\lambda_j := 1/\omega_j$. Then
\begin{equation}
\label{3.11} \frac{\partial^2 \overline{H}_4}{\partial a_j
\partial a_k} = \frac{3 \beta}{2 N}
\begin{pmatrix}
\frac{3}{4}\lambda_1^2 & \lambda_1 \lambda_2  &  & & \lambda_1 \lambda_N \\
\lambda_2 \lambda_1     & \frac{3}{4}\lambda_2^2 &  & & \lambda_2 \lambda_N \\
\vdots                 &          &              \ddots & & \vdots \\
\lambda_{N-1} \lambda_1 & \lambda_{N-1} \lambda_2 & & \frac{3}{4}\lambda_{N-1}^2 & \lambda_{N-1} \lambda_N\\
\lambda_N \lambda_1    & \lambda_N \lambda_2    &  & & \frac{1}{2}
\lambda_N ^2
\end{pmatrix}
\end{equation}
\begin{equation}
\label{3.12} \frac{\partial^2 \overline{H}_4}{\partial b_j
\partial b_k} = -\frac{3 \beta}{4 N}
\begin{pmatrix}
\lambda_1 ^2 & & & \\
 & \lambda_2 ^2 & & \\
 & & \ddots &                \\
 & & & \lambda_{\frac{N-1}{2}} ^2
\end{pmatrix}
\end{equation}
Clearly $\frac{\partial^2 \overline{H}_4}{\partial b_j \partial
b_k}$ is nondegenerate. After some algebra, one can check that
$$
\det \left(\frac{\partial^2 \overline{H}_4}{\partial a_j \partial
a_k} \right) = \left(\frac{3 \beta}{2 N}\right)^N \frac{2 N
-1}{2^N} \prod\limits_{j=1} ^N \lambda_j ^2,
$$
i.e., it is also nondegenerate.

 $ \hfill \square$

 Thus, the periodic KG lattice (\ref{1.2}) with  an odd number of particles can, after normalization, be viewed as a perturbation
 of a nondegenerate integrable Hamiltonian system, namely its fourth order normal form. Therefore, by the KAM theorem
 almost all low-energy solutions of (\ref{1.2}) are periodic or quasi-periodic and live on invariant tori.

\begin{Cor}
\label{Cor2} When $N$ is even, the truncated normal form
$\overline{H} = H_2 + \overline{H}_4$ of the periodic KG lattice
is Liouville integrable with the quadratic integrals $a_k, \, 1
\leq k \leq \frac{N}{2}$ and $a_N$, $b_k - b_{\frac{N}{2}-k}, \, 1
\leq k < \frac{N}{4}$ and $c_{\frac{N}{4}}$ (if $\frac{N}{4} \in
\mathbb{N}$) and the quartics
\begin{equation}
\label{3.15} \mathcal{K}_k = 3 \frac{c_k c_{\frac{N}{2}-k} - d_k
d_{\frac{N}{2}-k}}{\omega_k \omega_{\frac{N}{2}-k}} - \frac{3}{4}
\left(\frac{b_k ^2}{\omega_k ^2} + \frac{b_{\frac{N}{2}-k}
^2}{\omega_{\frac{N}{2}-k} ^2} \right), \quad 1 \leq k <
\frac{N}{4}.
\end{equation}
\end{Cor}
\noindent {\bf Proof.} This follows from simple calculations of
all Poisson brackets using (\ref{brackets}).

Let us deal with the exceptional cases. First, in the case of
$N=2$ particles the frequencies are incommensurable, i.e., there
are no resonances. From (\ref{h4}) we obtain
$$
\overline{H} = \omega_1 a_1 + \omega_2 a_2 + \frac{\beta}{4}
\left\{ \frac{3}{2} \left(\frac{a_1^2}{\omega_1^2} +
\frac{a_2^2}{\omega_2^2} \right)+ 6 \frac{a_1 a_2}{\omega_1
\omega_2} \right\},
$$
where $a_k = \frac{1}{2} (Q_k^2 + P_k^2), \, k= 1, 2$. The action
variables $a_k$ can be extended to global action-angle variables
$(a_k, \varphi_k)$ usually called symplectic polar coordinates.
Then the KAM condition is immediate.

Further, for the case with $N=4$ particles, we get from (\ref{h4})
and (\ref{relHopfv})
$$
\overline{H}_4 = \frac{\beta}{8} \left\{ \frac{3}{2}
\left(\frac{a_2^2}{\omega_2^2} + \frac{a_4^2}{\omega_4^2} \right)+
6 \frac{a_2 a_4}{\omega_2 \omega_4} + 6 \frac{a_1}{\omega_1}
\left(\frac{a_2}{\omega_2}+\frac{a_4}{\omega_4} \right) +
\frac{3}{2 \omega_1^2} (a_1^2 + c_1^2) \right\}.
$$
Denote $I_k = \frac{1}{2} (Q_k^2 + P_k^2), \, k= 1, \ldots, 4$.
Note that $a_1 = I_1 + I_3, \, c_1 = I_1 - I_3, \, a_2 = I_2, \,
a_4 = I_4$. Then $\overline{H}_4$ becomes
$$
\overline{H}_4 = \frac{\beta}{8} \left\{ \frac{3}{2}
\left(\frac{I_2^2}{\omega_2^2} + \frac{I_4^2}{\omega_4^2} \right)+
6 \frac{I_2 I_4}{\omega_2 \omega_4} + 6 \frac{I_1 + I_3}{\omega_1}
\left(\frac{I_2}{\omega_2}+\frac{I_4}{\omega_4} \right) +
\frac{3}{\omega_1^2} (I_1^2 + I_3^2) \right\}.
$$
The actions $I_k$ can be extended to global action-angle variables
$(I_k, \varphi_k)$ (symplectic polar coordinates) and the KAM
condition is straightforward (compare with \cite{OC2}).
 $ \hfill \square$

{\bf Proof of Theorem \ref{th2}.} Part (i) is proved by Corollary
1, whereas part (ii) comes after Corollary 2. $ \hfill
\blacksquare$

\section{KG lattice with fixed endpoints}

In this section we consider the KG lattice with  $n$ ($n \geq 3$,
not necessarily even) particles and with fixed endpoints
 \begin{equation}
 \label{5.1}
 q_0 = q_{n+1} = p_0 = p_{n+1} = 0.
\end{equation}
It was realized in \cite{Rink3} that such the FPU lattice with the
fixed boundary conditions can be viewed as an invariant subsystem
of the periodic FPU lattice with $N = 2 n + 2 $ particles. This
invariant subsystem is obtained by the fixed point set of the
compact group $\langle S\rangle = \{Id,S \}$. Since $S$ is also a
symmetry for the periodic KG Hamiltonian this constriction is
applicable here and we will describe it briefly.

Define the set
\begin{equation}
\label{5.2}
 Fix \langle S \rangle := \{ (q, p) \in T^* \mathbb{R}^N | S (q, p) = (q, p) \} .
 \end{equation}
By the definition of $S$ (\ref{symS}) we get
$$
q_j = - q_{2n+2 -j}, \quad p_j = - p_{2n+2 -j} \quad \forall j
$$
from where it follows that $q_0 = q_{n+1} = q_0 = q_{n+1} = 0$.
$Fix \langle S \rangle$ is a symplectic manifold with symplectic
form obtained from $dq \wedge dp$ with restriction on $Fix \langle
S \rangle$. One can takes as coordinates on $Fix \langle S
\rangle$ $(q_1, \ldots, q_n, p_1,\ldots, p_n)$. Then
 the lattice with fixed endpoints and $n$ particles is described on $Fix \langle S \rangle$ by
 the Hamiltonian $H_{Fix \langle S \rangle}$.

Let us consider the periodic KG Hamiltonian with $N=2n+2$
particles. From the definition of the phonon particles
(\ref{obr}),
 we can obtain that $S$ acts on them in the following way
\begin{align}
\label{5.5}
S : & (Q_1, \ldots, -Q_{n+1}, Q_{n+2}, \ldots, Q_{2n+1}, -Q_{2n+2}; P_1,\ldots, P_{n+1}, P_{n+2}, \ldots, P_{2n+1}, P_{2n+2})  \\
\mapsto & (-Q_1, \ldots, -Q_{n+1}, Q_{n+2}, \ldots, Q_{2n+1},
Q_{2n+2}; -P_1, \ldots, -P_{n+1}, P_{n+2}, \ldots, P_{2n+1},-
P_{2n+2}). \nonumber
\end{align}
Hence,
\begin{equation}
\label{5.6}
 Fix \langle S \rangle := \{ (Q, P) \in T^* \mathbb{R}^N | Q_k = P_k = 0, \, 1 \leq k \leq n+1, \, Q_{2n+2}=P_{2n+2} = 0 \} ,
\end{equation}
which is a symplectic manifold isomorphic to $T^*\mathbb{R}^n$. We
can take as coordinates on $Fix \langle S \rangle$ $(x_k, y_k) =
(Q_{2n+2-k}, P_{2n+2-k}), \, k=1, \ldots, n$.

We have already constructed an $S$-invariant truncated normal form
$\overline{H}$ for the periodic KG lattice. It was realized in
\cite{Rink3} that in order to obtain the normal form of the
Hamiltonian $H_{Fix \langle S \rangle}$, one needs to restrict the
symmetric normal form $\overline{H}$ to $Fix \langle S \rangle$,
that is,
$$
\overline{H_{Fix \langle S \rangle}} = \overline{H}_{Fix \langle S
\rangle}.
$$
Clearly, $b_k = d_k = 0, \, 1 \leq k \leq n, \, a_N =
a_{\frac{N}{2}}=0$ and $c_k = - a_k, \, 1 \leq k \leq n$.
Introduce
\begin{equation}
\label{5.7} I_k := a_k = \frac{1}{2} (x_k ^2 + y_k ^2), \quad 1
\leq k \leq n .
\end{equation}
Then we have
\begin{thm}
\label{nfDbC} The KG lattice with $n$ particles and fixed boundary
conditions has the fourth order normal form $\overline{H} = H_2 +
\overline{H}_4$, where $ H_2 = \sum_{k=1} ^n \omega_k I_k, \quad
\omega_k := \sqrt{1 + 4 \sin^2 \frac{k \pi}{2 n +2}} $ and
\begin{equation}
\label{fixedbnf} \overline{H}_4 = \frac{\beta}{2 (2 n + 2)} \Bigg
\{ \frac{9}{4} \sum\limits_{1\leq k \leq n}
\frac{I_k^2}{\omega_k^2}
 + 6 \sum\limits_{1\leq k < l \leq n} \frac{I_k I_l}{\omega_k \omega_l}
 + 3 \sum\limits_{1\leq k < \frac{n+1}{2}}
\frac{I_k I_{n+1-k}}{\omega_k \omega_{n+1-k}} + \frac{3}{4}
\frac{I_{\frac{n+1}{2}}^2}{\omega_{\frac{n+1}{2}}^2} \Bigg \} .
\end{equation}
In the above formula the term with subscripts $\frac{n+1}{2}$
appears if $n$ is odd.

This normal form is completely integrable with the quadratic first
integrals $I_k, \, k = 1, \ldots, n$ and it is KAM nondegenerate.
\end{thm}

\noindent {\bf Proof.} It follows from Theorem \ref{thNF} and the
explanations above that the quantities $I_k$ are Poisson
commuting, so the complete integrability is clear.
 The variables $I_k$ can be extended to global action-angle variables $(I_k, \varphi_k)$ - symplectic polar coordinates.
 It remains to verify the nondegeneracy of the normal form $\overline{H}$. Denote the Hessian of $\overline{H}$ with
 $\mathcal{H}$ and let $\lambda_j = 1/\omega_j$ as before. We have
\begin{equation}
\label{5.8} \mathcal{H} = \frac{\beta}{2(2n+2)} \triangle_n
\frac{3}{2} F_n \triangle_n,
\end{equation}
where $\triangle_n = diag (\lambda_1, \ldots, \lambda_n)$ and
$F_n$ is an $n \times n$ matrix which for $n$ even, respectively
odd takes the form
$$
\underbrace{
\begin{pmatrix}
3      & 4 & \ldots &            &          &\ldots  & 4 & 6 \\
4      & 3 & 4      & \ldots     & \ldots   &   4    & 6 & 4  \\
\vdots &   & \ddots &            &          & \iddots&   & \vdots \\
       &   &        &         3 &   6      &            &        \\
       &   &        &          6 &   3      &            &        \\
\vdots &  & \iddots &           &          & \ddots &   & \vdots  \\
4      & 6 & 4      & \ldots     & \ldots   &   4    & 3 & 4  \\
6      & 4 & \ldots &           &          &\ldots  & 4 & 3
\end{pmatrix}
}_{\mbox{ $n$ even}}, \quad \underbrace{\begin{pmatrix}
3      & 4 & \ldots &          &  &          &\ldots  & 4 & 6 \\
4      & 3 & 4      & \ldots   &  & \ldots   &   4    & 6 & 4  \\
\vdots &   & \ddots &          &  &          & \iddots&   & \vdots \\
       &   &        &         3& 4&   6      &        &   &        \\
  4    & 4 &        &         4& 4&   4      &        & 4 & 4      \\
       &   &        &         6& 4&   3      &        &   &        \\
\vdots &   & \iddots&          &  &          & \ddots &   & \vdots  \\
4      & 6 & 4      & \ldots   &  & \ldots   &   4    & 3 & 4  \\
6      & 4 & \ldots &          &  &          &\ldots  & 4 & 3
\end{pmatrix}}_{\mbox{ $n$ odd}}.
$$
Then
$$
\det (\mathcal{H}) = \left(\frac{\beta}{2(2n+2)}\right)^n
\prod\limits_{j=1} ^n \lambda_j ^2  \left(\frac{3}{2}\right)^n
\det(F_n).
$$
After some linear algebra $F_n$ is shown to be nonsingular, from
where the nondegeneracy of $\overline{H}$ follows.

$ \hfill \blacksquare$

As a consequence of the above result, we conclude that almost all
low-energy solutions  of the KG lattice with fixed endpoints
$H_{Fix \langle S \rangle}$ are quasi-periodic and live on
invariant tori.

\section{Concluding Remarks}

In present paper, we deal with the integrability of the KG
lattices. First, we study the periodic KG lattice with $N$
particles (\ref{1.2}) and quartic potential
$$
V (x) =  \frac{a}{2} x^2 + \frac{\beta}{4} x^4.
$$
We have shown that this Hamiltonian system is integrable in
Liouville sense only when $\beta = 0$. For this we use
Differential Galois theory and Morales-Ramis approach.

The considered system enjoys the same important discrete
symmetries $R, S$ (\ref{symT}), (\ref{symS}) as in the periodic
FPU chain. Following \cite{Rink1} we construct an $R,S$-symmetric
resonant forth order normal form $\overline{H}$. This normal form
happens to be Liouville integrable. It is  similar to the normal
form of the periodic FPU $\beta$-chain, but it is natural to be
expected. Hence, the periodic KG lattice can be considered as a
perturbation of its integrable Bikhoff normal form.

If $N$ is odd the integrals of the normal form are quadratic. The
global action-angle variables can be introduced and it turns out
that this normal form is KAM nondegenerated. This proves the
existence of many quasi-periodic solutions in the dynamics of the
periodic KG lattice at low-energy level.

The resonant normal form with  $N$ even admits certain set of
quartic integrals in addition to the quadratic ones. Probably, it
would be interesting to explore the geometry of the system,
defined by  this normal form. One can assume that the things are
similar to the even periodic FPU chain \cite{Rink2} up to small
modifications due to the extra degree of freedom.

Next, we consider the KG lattice with fixed endpoints. Such a
system can be considered as an invariant symplectic submanifold of
a larger periodic KG lattice. For this the discrete symmetry $S$
is utilized. Then the normal form in this case is easy to get from
the previous result, and hence, it is integrable. Further, KAM
theorem applies which implies that almost all low-energetic
solutions are quasi-periodic.

Finally, we  notice that the results of this paper do not provide
an answer to one of the most important problems: what happens in
the dynamics of the KG lattice when the number of particle becomes
larger and larger.

Let us emphasize again the importance of discrete symmetries, and
in particular, the symmetry  $S$ in the carrying out of the above
analysis. This leads to the following question: what happens when
we drop the symmetry $S$? We can ask the same thing in a different
way: Can the results of this paper be extended for the KG lattice
with the potential
$$
V (x) =  \frac{a}{2} x^2 + \alpha x^3 + \frac{\beta}{4} x^4, \quad
\alpha \neq 0,
$$
which is more relevant in studying the the dynamic of
low-energetic solutions in the DNA model?

It is clear that  the non-integrability result of Theorem
\ref{th1} can easily be extended in the same lines as in the
Appendix. The formal computation of the normal form would be more
difficult, because we have to transform away the third order
terms. However, straightforward calculations for the
low-dimensional periodic KG lattices  with $a=1$ and $N=2, \ldots,
6$ show that resonant third order terms do not appear in the
corresponding normal forms (see \cite{OC2}). Hence, these normal
forms remain integrable for the latter potential. In view of the
applications to the DNA models, it is clearly of some interest to
calculate these normal forms in the general case.

\vspace{3ex}

{\bf Acknowledgements.} This work is partially supported by grant
DN 02-5 of Bulgarian Fond "Scientific Research".

\appendix

\section{Proof of Theorem \ref{th1}}

The proof of Theorem \ref{th1} is based on
Ziglin-Morales-Ruiz-Ramis theory. The main result of this theory
merely says that if a Hamiltonian system is completely integrable
then the identity component of the Galois group of the variational
equation along certain particular solution is abelian. The
necessary facts and results about differential Galois theory and
its relations with the integrability of Hamiltonian systems, can
be found in \cite{M,MRS1,MR2,SvP}.

In the applications if one finds out that the identity component
of the Galois group is non-commutative, then this implies
non-integrability. However, if this component turns out to be
abelian, one needs additional steps to prove non-integrability as
it is carried out below.

It is also assumed that throughout this appendix all variables are
complex: $t \in \mathbb{C}, q_j \in \mathbb{C}, p_j \in
\mathbb{C}, j = 1, \ldots, N$, but we keep the parameters $a,
\beta$ real. The proof goes in the following lines. We obtain a
particular solution and write the variational equation along this
solution. It appears that the identity component of its
differential Galois group is abelian. In order to obtain an
obstacle to the integrability, we study the higher variational
equations. Although higher variational equations are not actually
homogeneous equations, they can be put in the differential Galois
context. Their differential Galois groups are in principle
solvable. One possible way to show that some of them is not
abelian is to find a logarithmic term in the corresponding
solution (see \cite{MRS1,MR2}). We obtain such a logarithmic term
in the solution of the second variational equation when $\beta
\neq 0$. Then the non-integrability of the Hamiltonian system
follows.

\vspace{2ex}

\noindent {\bf Proof.} Suppose $\beta \neq 0$.  The following
Proposition is immediate.
\begin{prop}
\label{prop1} The Hamiltonian system corresponding to (\ref{1.2})
admits a particular solution
\begin{equation}
\label{a1} q_j (t) := q  = sn (\sqrt{a + \beta/2} \, t, \kappa),
\quad p_j (t) = \frac{d}{d t} q_j (t), j = 1, \ldots, N ,
\end{equation}
where $sn$ is the Jacobi elliptic function with the module $\kappa
= \sqrt{\frac{-\beta/2}{ a + \beta/2}}$.
\end{prop}
 $ \hfill \square$

\noindent {\bf Remark 3.} It is assumed that $\beta > 0$ which is
not restrictive. In any case, the solution is expressed via Jacobi
elliptic functions and one can proceed in the same way.

It is straightforward that $T_1 = \frac{4 K}{\sqrt{a + \beta/2}}$
and $T_2 = \frac{2 i K'}{\sqrt{a  + \beta/2}}$ are the periods of
(\ref{a1}). Here $K, K'$ are the complete elliptic integrals of
the first kind. In the parallelogram of the periods, the solution
(\ref{a1}) has two simple poles
\begin{equation}
\label{a2} t_1 = \frac{iK'}{\sqrt{a  + \beta/2}}, \quad t_2 =
\frac{2 K + i K'}{\sqrt{a + \beta/2}} .
\end{equation}
From the expansion of the $sn$ in the neighborhood of the pole
$t_1$ we get
\begin{equation}
\label{a3} q  = \frac{1}{\sqrt{-\beta/2}}\Big[\frac{1}{t-t_1} +
\frac{4+a}{6}(t-t_1) + \gamma_3 (t-t_1)^3 + \gamma_5 (t-t_1)^5 +
\ldots \Big],
\end{equation}
where $\gamma_3$ is an arbitrary constant and $\gamma_5$ amounts
to
\begin{equation}
\label{a4} \gamma_5 = \frac{1}{14} \Big[ \frac{16}{27} + 4
\gamma_3 + \frac{4a}{9} + a\gamma_3 + \frac{a^2}{9} +
\frac{a^3}{108}\Big].
\end{equation}

Denoting by $\xi^{(1)}_j = d q_j, \eta^{(1)}_j = d p_j, j=1,2$ the
variational equations (VE) (written as second order equations) are
\begin{equation}
\label{a5} \ddot{\xi}^{(1)}_j + (2 + V'' (q)) \xi^{(1)}_j -
\xi^{(1)}_{j+1} - \xi^{(1)}_{j-1} = 0, \qquad j = 1, \ldots, N.
\end{equation}
\begin{prop}
\label{prop2} The identity component of the differential Galois
group of (VE) (\ref{a5}) is abelian.
\end{prop}
{\bf Proof.} To see this we first denote
\begin{equation}
\label{a6} \xi^{(1)} :=
\begin{pmatrix}
\xi^{(1)}_1 \\
\xi^{(1)}_2 \\
\vdots \\
\vdots \\
\xi^{(1)}_{N-1} \\
\xi^{(1)}_N
\end{pmatrix}, \quad
K_N (q) :=
\begin{pmatrix}
2+V'' (q) & -1 &   &   & -1 \\
-1 & 2+V''(q) & -1  &   &  \\
\\
   & \ddots & \ddots   & \ddots       &  \\
   \\
  &   &-1   &  2+V''(q) & -1   \\
-1 &   &   & -1 & 2+V''(q)
\end{pmatrix} .
\end{equation}
Then (VE) can be written as
\begin{equation}
\label{a7} \ddot{\xi}^{(1)} + K_N (q) \, \xi^{(1)} = 0.
\end{equation}
The structure of the matrix $K_N (q)$ is similar to that of $L_N$
in (\ref{2.6}). This suggest using the  linear transform
\begin{equation}
\label{a8} \xi^{(1)} = M y^{(1)}
\end{equation}
with already defined matrix $M$, which decouples the system
(\ref{a7})
\begin{equation}
\label{a9} \ddot{y}^{(1)} + D_N (q) \, y^{(1)} = 0,
\end{equation}
where $D_N (q):= M^{-1} K_N (q) \, M = diag (\Lambda_1 ^2, \ldots,
\Lambda_N ^2)$, with $\Lambda_j ^2 := V'' (q) + 4 \sin^2 \frac{\pi
j}{N}, \, j = 1 , \ldots, N$. In coordinate form the above system
can be written as
\begin{equation}
\label{a10} \ddot{y}_j ^{(1)} + [a + 4 \sin^2 \frac{\pi j}{N} + 3
\beta sn^2 (\sqrt{a + \beta/2} \, t, \kappa) ] y_j ^{(1)} = 0,
\quad j = 1, \ldots, N.
\end{equation}
After changing the independent variable $\tau := \sqrt{a +
\beta/2} \, t, \, ' = d /d \tau$ we can see that each of these
equations is a Lam\'{e} equation in Jacobi form.
\begin{equation}
\label{a11} \left(y_j ^{(1)}\right)''  = \Big [(\tilde{n}+1) \,
\tilde{n} \, \kappa^2 sn^2 (\tau, \kappa) - \frac{a+4 \sin^2
\frac{\pi j}{N} }{a + \beta/2} \Big] y_j ^{(1)},
\end{equation}
with $\tilde{n} = 2$. The identity component of the Galois group
of such equations is known to be  isomorphic to $
\begin{pmatrix}
1 & 0 \\
\nu_j & 1
\end{pmatrix}, \nu_j \in \mathbb{C}
$. Therefore, the identity component $G^0$ of the Galois group of
(VE) is represented by the matrix group
$$
 G^0 = \left \{ \begin{pmatrix}

 \begin{matrix}
 1 & 0 \\
 \nu_1 & 1
 \end{matrix} & &   &  \\
& \ddots &  &\\
  & &
 \begin{matrix}1 & 0 \\
 \nu_{N} & 1
 \end{matrix} &
  \end{pmatrix}, \quad
  \nu_j \in \mathbb{C}, \, j = 1, \ldots, N \right \}
  $$
  and it is clearly abelian.

  $ \hfill \square$

Now, let us consider the higher variational equations along the
particular solution (\ref{a1}). We write
\begin{equation}
\label{a13} q_j =  q + \varepsilon \xi^{(1)} _j + \varepsilon^2
\xi^{(2)} _j + \varepsilon^3 \xi^{(3)} _j + \ldots, \quad p_j =
\dot{q}_j ,
\end{equation}
where $\varepsilon$ is a formal parameter and substitute these
expressions into the Hamiltonian system governed by (\ref{1.2}).
Comparing the terms with the same order in $\varepsilon$ we obtain
consequently the variational equations up to any order.

The first variational equation is, of course, (\ref{a5})
$(\mathrm{VE}_1) = (\mathrm{VE})$. For the second variational
equation we have
\begin{equation}
\label{a14} \ddot{\xi}^{(2)} _j + (2 + V'' (q)) \xi^{(2)} _j -
\xi^{(2)} _{j+1} - \xi^{(2)}_{j-1} = -\frac{1}{2} V''' (q) (
\xi^{(1)} _j)^2 , \qquad j = 1, \ldots, N.
\end{equation}
Denote
\begin{equation}
\label{a15} \xi^{(2)} := (\xi^{(2)} _1, \ldots, \xi^{(2)} _N)^T,
\qquad f^{(2)} := - 3 \beta q \, (( \xi^{(1)} _1)^2, \ldots, (
\xi^{(1)} _N)^2 )^T.
\end{equation}
Then the second variational equation can be written as
\begin{equation}
\label{a16} \ddot{\xi}^{(2)} + K_N (q) \, \xi^{(2)} = f^{(2)}.
\end{equation}
In this way we can obtain a chain of linear non-homogeneous
differential equations
\begin{equation}
\label{a17} \ddot{\xi}^{(k)} + K_N (q) \,  \xi^{(k)} = f^{(k)}
(\xi^{(1)}, \ldots, \xi^{(k-1)}), \quad k = 1, 2, \ldots ,
\end{equation}
where  $f^{(1)}=0$. The above equation is usually called $k$-th
variational equation $(\mathrm{VE}_k)$ (here is is written in the
form of second-order equation). As it was mentioned above higher
variational equations are solvable. Indeed, if $\Xi (t)$ is a
fundamental matrix of $(\mathrm{VE}_1)$, then the solutions of
$(\mathrm{VE}_k), k > 1$  can be found by quadratures
\begin{equation}
\label{a18} \xi^{(k)} = \Xi (t) r_k ,
\end{equation}
where
\begin{equation}
\label{a19} \dot{r}_k = \Xi^{-1} (t) f^{(k)}.
\end{equation}

Let us study the local solutions of $(\mathrm{VE}_2)$. We make the
linear change (with above introduced matrix $M$)
$$
\xi^{(2)} = M y^{(2)}.
$$
The  system (\ref{a16}) becomes
\begin{equation}
\label{a20} \ddot{y}^{(2)} + D_N (q) y^{(2)} = \tilde{f}^{(2)},
\end{equation}
where
$$
\tilde{f}^{(2)} := M^{-1} f^{(2)},
$$
that is, in these coordinates $(\mathrm{VE}_2)$ decomposes into
$N$ second order differential equations. Therefore, it is enough
to show that the identity component of the Galois group of one of
them is non-commutative, which implies non-commutativity of the
Galois group of $(\mathrm{VE}_2)$, and hence, non-integrability of
the Hamiltonian system under consideration.

To keep the things simple, we take as a solution of
$(\mathrm{VE}_1)$ (\ref{a10})
\begin{equation}
\label{sol} y^{(1)} _1 = y^{(1)} _2 = \ldots = y^{(1)} _{N-1} = 0
\end{equation}
and $y^{(1)} _N \neq 0$ satisfying
\begin{equation}
\label{a21} \ddot{y}^{(1)} _N  + [ a  + 3 \beta (sn (\sqrt{a +
\beta/2} \, t, \kappa))^2 ] y^{(1)}_N = 0.
\end{equation}
In what follows we need the expansions around $t_1$ of the
fundamental system of the solutions with unit Wronskian for
(\ref{a21}). We get
\begin{align}
\label{a22}
y^{(1)} _{N,1} & =  \frac{1}{(t-t_1)^2} - \frac{8+a}{6} - \frac{a+6-9\gamma_3}{3}(t-t_1)^2 \nonumber \\
 & + \left(\frac{56+11a}{27} - \frac{(91+9a)\gamma_3}{18} - \frac{a^2}{54} - \frac{a^3}{216} \right)(t-t_1)^4 + \ldots   \\
y^{(1)} _{N,2} & =  \frac{1}{5}(t-t_1)^3 + \ldots . \nonumber
\end{align}
Then the fundamental matrix $\Xi_N (t)$ of (\ref{a21}) can be
written as
\begin{equation}
\label{a23} \Xi_N (t) =
\begin{pmatrix}
y^{(1)} _{N,1} & y^{(1)} _{N,2} \\[2ex]
\dot{y} ^{(1)} _{N,1} & \dot{y} ^{(1)} _{N,2}
\end{pmatrix}
\qquad \Xi_N ^{-1} (t) =
\begin{pmatrix}
\dot{y}^{(1)} _{N,2} & -y^{(1)} _{N,2} \\[2ex]
-\dot{y} ^{(1)} _{N,1} & y ^{(1)} _{N,1}
\end{pmatrix} .
\end{equation}

Now, we study the local solutions around $t_1$ only of the last
equation in the system (\ref{a20}) taking into account
(\ref{sol}), (\ref{a21}), (\ref{prava}) and (\ref{obr})
\begin{equation}
\label{a24} \ddot{y}^{(2)} _N + [ a  + 3 \beta (sn (\sqrt{a +
\beta/2} \, t, \kappa))^2 ] y^{(2)}_N = - \frac{3 \beta}{\sqrt{N}}
\, q \, (y^{(1)}_N)^2 .
\end{equation}

In this case (\ref{a19}) becomes
$$
\frac{d}{d t}
\begin{pmatrix}
r_{21} \\ r_{22}
\end{pmatrix} =
 \Xi_N ^{-1} (t)
 \begin{pmatrix}
 0 \\
- \frac{3 \beta}{\sqrt{N}} \, q \, (y^{(1)}_N)^2
\end{pmatrix}
  =
  \begin{pmatrix}
   y^{(1)}_{N,2} \frac{3 \beta}{\sqrt{N}} \, q \, (y^{(1)}_N)^2 \\
 - y^{(1)}_{N,1} \frac{3 \beta}{\sqrt{N}} \, q \, (y^{(1)}_N)^2
\end{pmatrix}
$$
We are looking for a component of above vector with a nonzero
residuum at $t = t_1$. This would imply the appearance of a
logarithmic term. After some calculations making use of (\ref{a3})
and (\ref{a22}), the residue at $t = t_1$ of the second component
$- y^{(1)}_{N,1} \frac{3 \beta}{\sqrt{N}} \, q \, (y^{(1)}_N)^2 $
with the specific representatives turns out to be
\begin{equation}
\label{residue} \mathrm{Res}_{|_{t=t_1}} - 3
\frac{\beta}{\sqrt{N}} \, q \,  y^{(1)}_{N,1} \, (y^{(1)}_{N,1}
)^2 = \frac{3\beta}{\sqrt{-\beta/2}}\Big[\frac{a^3}{252} +
\frac{a^2}{21} + \frac{4a}{21} + \frac{3a\gamma_3}{7} +
\frac{73\gamma_3+16}{63} \Big].
\end{equation}
Since $\gamma_3$ is an arbitrary parameter, we choose it in such a
way that the expression in the square brackets of (\ref{residue})
does not vanish for $a > 0$. There are many such values of
$\gamma_3$, for instance, we can take $\gamma_3 = 1$. Recall that
by assumption $\beta \neq 0$. We have obtained a nonzero residuum
at $t = t_1$, which implies the appearance of a logarithmic term
in the solutions of $(\mathrm{VE}_2)$. Then its  Galois group is
solvable but not abelian. Hence, we conclude the non-integrability
of the Hamiltonian system (\ref{1.2}).

$\hfill \blacksquare$

\end{document}